# Promoting Instructional Change via Co-Teaching

Charles Henderson
*Physics Department and Mallinson Institute for Science Education, Western Michigan University*

Andrea Beach
*Department of Teaching, Learning and Leadership, Western Michigan University*

Michael Famiano
*Physics Department, Western Michigan University*

**Abstract**

Physics Education Research (PER) has made significant progress in developing knowledge about teaching and learning as well as effective instructional strategies based on this knowledge. Disseminating knowledge and strategies to other faculty, however, has proven difficult. Co-teaching is a promising and cost-effective alternative to traditional professional development that may be applicable in many situations. In this article, we discuss the theoretical background of co-teaching and describe our initial experience with co-teaching. A new instructor (MF) co-taught with an instructor experienced in PER-based reforms (CH). The pair worked within the scaffolding of the course structure typically used by the experienced instructor and met regularly to discuss instructional decisions. An outsider (AB) conducted separate interviews with each instructor at the beginning, middle, and end of the semester and observed several class sessions. Classroom observations show an immediate use of PER-based instructional practices by the new instructor. Interviews show a significant shift in the new instructor's beliefs about teaching and intentions towards future use of the PER-based instructional approaches.

# 1. Introduction

Many funding models for promoting educational reform in the teaching of college science courses are based on the premise that an individual or group of faculty develops and tests a research-based instructional strategy and then disseminates this new strategy to colleagues at their home institution and at other institutions. Dissemination typically takes the form of large scale transmission-oriented activities such talks, workshops, and publications. This model seems to be based on the implicit assumption that if the new method is broadcast to enough instructors, some will likely become users. There is little evidence to support this model of reform, and, in fact, it appears quite common for new strategies to fade away even at the home institution once funding runs out and the original developers move on to other things [1].

In this paper we identify some of the difficulties with common transmissionist dissemination models and describe an alternative model that may be helpful in overcoming some of these difficulties.

# 2. Theoretical Background

In this section we briefly describe several lines of research that have shaped our thinking about this project. We do not intend for this to be a comprehensive review of the vast literature on instructional change, but rather we seek to give the reader an understanding of our perspectives on several important themes within this literature.

## A. Fundamental Instructional Changes Are Rare

Most reforms in science teaching call for a significant shift in the role of teachers in the educational system [1-3]. In "traditional" educational systems, the role of the teacher is to be a content expert who can impart his or her knowledge to students [4]. Most reforms call for a shift in this role to the teacher as a facilitator of the learning process. This is true both at the K-12 level where the National Science Education Standards call for instructional environments where "students actively develop their understanding of science by combining scientific knowledge with reasoning and thinking skills" (ref [2], p. 2) and at the college level, where the National Research Council's Committee on Undergraduate Science Education calls for undergraduate teaching faculty to "be prepared to use combinations of inquiry-based, problem-solving, information-gathering, and didactic forms of instruction under appropriate classroom circumstances that promote conceptual understanding and students' ability to apply knowledge in new situations" (ref [5], p. 27).

Larry Cuban refers to such a shift in the role of teachers as a fundamental change and distinguishes between fundamental and incremental changes [6]. "*Incremental changes* aim to improve the efficiency and effectiveness of existing structures, cultures, and processes. The premise behind incremental change is that the basic structures are sound but need improving to remove defects that hinder effectiveness and efficiency." (ref [6], p. 63) "*Fundamental changes* are those that aim to alter drastically the core beliefs, behaviors, and structures. . . .[The premise is that] the basic structures, cultures, and processes are flawed at the core and need a complete overhaul, not simple renovations." (ref [6], p. 64) This distinction between fundamental and incremental changes is important because fundamental changes always face significant resistance while incremental changes often do not [7]. Thus, it is likely that different techniques are needed to bring about each type of change.

*Fundamental instructional changes conflict with the existing culture*

Research points out that a major obstacle to the implementation of a fundamental instructional change is that instructors attempting to change traditional practices are already acculturated into and surrounded by a culture that reflects their current practices [8-11]. Thus, they must undergo a difficult process of learning and cultural change. Even when instructors are able to successfully make internal changes they are typically still immersed in their current situation. Many aspects of this situation likely conflict with their new teaching culture [11]. For example, in an interview study, physics faculty cited strong situational constraints that made it difficult to teach in a non-traditional manner [12]. Commonalities such as large class sizes, broad content coverage expectations, classroom infrastructure, scheduling constraints, and poor student preparation/motivation all appear to favor traditional instruction. Another significant barrier to instructional change is that the current incentive structures at most universities place significant emphasis on excellence in disciplinary research, but little emphasis on excellence in teaching [1, 5, 6]. In addition, students often resist new instructional strategies [13]. In a traditional science class, students and instructors often abide by a "hidden contract" whereby students are responsible for sitting quietly and asking clarifying questions while teachers are responsible for presenting clear lectures and solving example exercises that are not too different from test questions [14]. When an instructor attempts to change this contract, many students feel threatened and resist [15].

*Instructional innovations require local customization*

Most instructors recognize that it is unlikely that any set of instructional materials, no matter how carefully developed, will match the constraints of their specific classroom situation. Therefore, even when instructors are convinced that there is an instructional problem (i.e., their existing practice does not achieve an instructional goal they value) which could potentially be solved by a change in their teaching practices; and even when they are convinced that a new practice is appropriate, they still must have the knowledge, skill, and confidence to customize the new practice to their own situation. Although the developers of instructional strategies may seek to improve the fidelity of implementation through their dissemination strategies (for example, by publishing materials that are difficult for instructors to modify) "there is no way to avoid the local reconstruction of the practice [instructional strategy], as local staff make sense of it in their own context" (ref [16], p. 33).

Because of the conflicts with existing cultures (both personal and institutional), new instruction that calls for fundamental changes (e.g., transforming classrooms from passive to active learning) is often changed by instructors and implemented as incremental changes. Although such implementation may keep some of the surface features of the innovation, it is essentially traditional instruction. For example, research suggests that instructors may implement Peer Instruction [15], but without the peer interaction component [17, 18]. This allows them to streamline their lecture and maintain a passive classroom, while appearing on the surface to have implemented a research-based innovation. Similarly, Boice reports that new college faculty self-described student-centered instruction was only confirmed by observations in about half of the cases, indicating that these faculty may have assimilated the ideas of student-centered instruction into their mental models of instruction such that some aspects that conflict with the existing cultures are modified to align with these cultures. Henderson and Dancy have developed the term *inappropriate assimilation* to describe this type of "adoption" [18]. Innovations requiring fundamental changes appear to be quite susceptible to inappropriate assimilation [19-23]. One unfortunate result of this phenomenon is that instructors who think that they have adopted an instructional strategy, but have really inappropriately assimilated it into their prior instructional practices, may conclude that the strategy is ineffective.

**B. Standard Dissemination Methods Are Expensive and Not Suitable for Promoting Fundamental Changes**

Common methods of dissemination include talks, papers, and workshops aimed at convincing individual faculty to change their instruction and giving them information and materials in support of a specific research-based strategy. As described above, for instructors to fundamentally change their instruction, they must simultaneously transform their personal views about teaching and learning to align with the new instruction as well as use this understanding to adapt the new instruction to their unique situation. Dangers along the way include numerous types of external resistance from situational factors and internal resistance from prior patterns of thinking and acting (which can result in inappropriate assimilation).

While standard dissemination strategies may be suitable for the promotion of incremental changes, it is unlikely that such a transmissionist approach to professional development can promote fundamental changes. This is for the same reasons that it is unlikely that transmissionist instructional strategies for science students can promote significant changes in student mental models. In addition, as Yerushalmi et. al. argue, most standard professional development does not seek to develop an understanding of the conceptual framework that underlies faculty instructional decisions [24]. This is similar to the traditional instructional practice of providing the same instruction to all students and all classes without developing an understanding of the students' initial knowledge and beliefs.

An additional barrier to standard transmissionist dissemination strategies is the complex nature of teaching itself. Similar to any complex task, much of a teacher's decision-making is implicit [25, 26]. It would be an overwhelming task for a curriculum developer to make all of the necessary implicit decisions explicit and equally overwhelming for an instructor to attempt to internalize these decisions. The ability to make "correct" decisions implicitly is learned through experience and reflection [26-29]. Of course, what is correct may be different in different instructional cultures. For example, from the perspective of traditional instruction, it may be very important to learn how to respond to student questions with clear and thorough explanations. From the perspective of many research-based instructional strategies it may be more important to learn how to respond to student questions with appropriate questions for the student. Thus, the development of this implicit knowledge is shaped by a larger organizing perspective [30]. An instructor who is in an instructional environment that supports traditional instruction and has personal beliefs consistent with traditional instruction will need substantial support to develop the implicit knowledge necessary to implement research-based instruction. It is likely that personal cultural changes, changes in implicit instructional knowledge, and changes in teaching practice occur simultaneously.

Although cost data is not commonly reported in the educational research literature, cost effectiveness of dissemination strategies is clearly an important consideration. In terms of traditional dissemination strategies, cost figures are available in an evaluation of the NSF Undergraduate Faculty Enhancement (UFE) program [31]. According to the evaluation, between 1991 and 1997, UFE made $60,963,917 in awards for faculty development workshops. This funded over 14,400 undergraduate faculty from all types of institutions to attend over 750 workshops. In telephone interviews conducted with 1,118 faculty participants in workshops during 1996 and 1997, 81% reported that they had made at least moderate changes to their own courses or developed new courses. Thus, extrapolating to all 14,400 faculty participants, the faculty development workshops cost, on average, $4200 per participant and each moderate or higher change cost $5,200. In addition, 40% of faculty participants reported making changes to teaching methods (other more common categories of changes included introducing new content – 67%, and introducing new lab techniques or technologies – 67%). Thus, each moderate or higher

change in teaching methods cost $10,600 – or $12,750 in 2004 dollars [32]. We consider this a low estimate for a fundamental change since not all changes in teaching methods are fundamental changes and, as discussed earlier, self-report data can be biased in terms of over reporting changes.

**C. Establishment of Teaching Styles by New Faculty**

Many new science faculty have held TA appointments in graduate school. However, few have actually taught a course of their own before their first faculty position. Thus, this is a formative time in the development of an instructor's teaching style and likely an ideal time for interventions aimed at promoting non-traditional instructional practices [33]. This is, for example, the philosophy behind the Physics New Faculty Workshops [34].

An alternative view is that for new faculty on the tenure-track any departure from traditional instruction is dangerous because such changes may require more time than traditional instruction and result in lower student ratings – especially at first [1]. Studies of new faculty, however, show that it is quite common for them to spend a majority of their time on instructional activities and receive poor student ratings under normal conditions. Robert Boice studied 77 new tenure-track faculty at two different universities (one with a research emphasis and one with a teaching emphasis) via interviews and observations during their first year [35]. By the middle of their first semester, most of the new faculty complained about the lack of collegial support and reported that lecture preparation dominated their time. Few of the faculty reported teaching skill as depending on anything other than their knowledge of content and clear, enthusiastic presentation. Most described their classes as standard facts-and-principles lecturing and many had no plans for improving their teaching. Boice concludes that new faculty typically teach cautiously, defensively, and tend to blame low student ratings on external factors (such as poor students, heavy teaching loads, and invalid rating systems). He suggests that new faculty would benefit from programs that helped them find ways to increase student participation while at the same time avoid over preparing facts.

Thus, we argue that since new faculty already struggle with learning how to teach, this is the time to assist them in developing an instructional style based on educational research. This is much more effective than the alternative of allowing new faculty to struggle on their own to develop an unsatisfactory traditional instructional style and then provide support to try and change that style after the instructor obtains tenure.

**D. Dissemination and learning theory: Co-Teaching Theoretical Basis**

Disseminating an instructional strategy means teaching an instructor how to act and/or think in new ways. Thus, it is fruitful to think about dissemination through the lens of instructional paradigms. Farnham-Diggory proposes three instructional paradigms [36]: behavior, development, and apprenticeship. Each paradigm is characterized by its view about how experts and novices differ and the mechanism by which novices become experts. In a behavior paradigm, "novices and experts are on the same scale(s)" (p. 464) and a novice is transformed by accumulating something. In terms of dissemination, this paradigm is evident in some professional development that seeks to expand the options available to faculty in their instructional "tool kit". The assumption is that experts and novices use basically the same criteria for determining whether a particular tool is appropriate or not. The difference is that the experts have more tools and so are more likely to have an appropriate tool available for any given instructional situation. This paradigm is likely appropriate for disseminating incremental changes. In a development paradigm, novices and experts have different personal theories or qualitative models. Novices become experts through the process of perturbation. In terms of dissemination, this paradigm is evident in some professional development that seeks to help faculty develop an entirely new way of thinking about teaching and learning by contradicting their assumed transmissionist personal theories and helping them develop a particular set of personal theories compatible with constructivism. In an apprenticeship paradigm, "novices and experts are from different worlds, and a novice gets to be an expert through the mechanism of acculturation into the world of the expert. Actual participation in this world is critical for two reasons: (a) much of the knowledge that the expert transmits to the novice is tacit, and (b) the knowledge often varies with context." [ref [36], p. 466] Dissemination activities in the apprenticeship paradigm are not commonly found described in the research literature.

Nonetheless, they likely occur when a faculty member takes a sabbatical or accepts a post-doc position at an institution with a well-developed instructional strategy. The faculty member or post-doc then learns how to implement this strategy by working with the developer and assisting with the actual delivery of instruction. Apprenticeship is also the instructional paradigm that fits most closely with co-teaching.

**E. What is Co-Teaching?**

The practice of co-teaching was developed by Roth and Tobin as an alternative to the standard student teaching practice associated with most K-12 teacher preparation programs [37, 38]. In standard student teaching, the student teacher typically first observes a number of the master teacher's classes and then the student teacher takes over the class on their own. Roth argues that student teachers do not often develop the tacit knowledge necessary to be good teachers under this arrangement [37]. During co-teaching the student teacher and master teacher share responsibility for all parts of the class. Student teachers "begin to develop a feel for what is right and what causes us to do what we do at the right moment" [ref [37], p. 774]. Although we are aware that co-teaching activities have occurred at the college level at other institutions, we are not aware of any other studies that have sought to document the results of such an arrangement.

**3. WMU Co-Teaching Project**

From an apprenticeship perspective, the goal of co-teaching in the current study is to acculturate MF into research-based physics instruction as embodied in the design principles developed and enacted by the WMU Physics Teachers Education Coalition (PhysTEC) faculty – see Table 1. As discussed earlier, the largely tacit and context-dependent nature of teacher decision-making means that learning to teach in a PhysTEC-compatible manner requires more than just talking about teaching – it requires direct experience in the practice of teaching. This is especially true since the culture (that is, the assumptions and norms) of PhysTEC teaching is very different from the culture of traditional teaching.

**A. Co-Teaching Activities and Context**

The co-teaching took place in the lecture portion of an introductory calculus-based physics course at Western Michigan University. The 4 credit course (Phys 2050: Mechanics and Heat) met each weekday for 50 minutes and contained about 70 students, mostly engineering majors, in a stadium-style lecture hall with fixed seating. CH and MF were both listed as the instructor of record for the course. In addition to the lecture, each student also participated in a weekly 2-hour laboratory session taught by TAs under the supervision of other department staff. The lab activities, however, had been redeveloped under the PhysTEC project to have a focus on developing and extending student understanding of the relevant physics topics.

The class structure had been developed by CH as part of the WMU PhysTEC project through consultation with his colleagues as well as the educational research literature. Ten important course components are described on Table 2. CH had successfully used a similar course structure in previous semesters for both Phys 2050 and the next course in the introductory sequence (Phys 2070: Electricity and Light).

There were five basic co-teaching activities. Each of these will be described briefly below and then considered from the perspective of a cognitive apprentice instructional framework.

(1) CH and MF alternate being in charge of class each week. Although both of the instructors were present during each class session, they alternated being "in charge" of the class on a weekly basis. The person in charge would typically preside over any whole-class discussions or presentations. Much of the class time was spent by students working in assigned small groups. During this time both instructors would circulate around the lecture hall and interact with groups. The first draft of weekly quizzes or exams were developed by the instructor in charge and shared with the other instructor for comment.

(2) Weekly meetings between CH and MF to reflect on previous week and discuss initial plans for coming week. Each Friday, CH and MF met for approximately 1 hour. During that time they talked about how things went during the past week and any difficulties that arose. The instructor in charge of the following week would then present their initial plans, which were discussed. In addition to this weekly scheduled meeting, CH and MF frequently had shorter discussions about the course at other times.

(3) Course structure set up by CH to support PhysTEC design principles. As discussed earlier, the course structure was set up by CH based on his previous successful teaching of the course. It was specifically designed so that much of class time would be used having students working together in small groups while discussing important physics ideas. Details of the course structure are shown in Table 2.

(4) MF had access to materials used by CH in previous offerings of the course. At the beginning of the semester, CH gave MF a CD with electronic copies of all the course activities and assignments used in the previous semester. MF typically used, with minor modifications, about half of these and developed the other half of the course activities himself.

(5) MF teaches course on his own during subsequent semester. MF taught the same class on his own during the following semester -- Spring 2006.

**B. Co-Teaching and Apprenticeship**

There are 6 basic aspects of a cognitive apprenticeship instructional model (from ref [39], p. 43):
  i. Modeling: expert performs a task so novice can observe.
  ii. Coaching: expert observes and facilitates while novice performs task.
  iii. Scaffolding: expert provides support to help the novice perform the task.
  iv. Articulation: expert encourages novice to verbalize their knowledge and thinking.
  v. Reflection: expert enables novice to compare their performance with others.
  vi. Exploration: expert invites novice to perform additional tasks with decreasing support.

Table 3 shows how each of the co-teaching activities match with the cognitive apprenticeship instructional model.

**C. Co-Teaching Reduces Risk**

As described earlier, new instructors are typically risk averse and afraid of making mistakes that may hurt their chances of getting tenure. In terms of teaching, this manifests itself in two ways. One is that they do not want to receive any complaints from students or low student evaluations. The other is that new faculty are often warned by more experienced faculty not to spend too much time on teaching. Any teaching innovation has the potential for increasing student complaints as well as giving other faculty the impression that the new faculty member is placing an inappropriate priority on teaching over research. Thus, any departure from traditional instruction must be made as risk-free as possible – in terms of both student satisfaction and time demands. Co-teaching, as enacted in this project, does this in two ways. First, it allows the experienced instructor to set up a course structure that is known to work in the particular context. This structure is further supported by modeling and coaching within the context. This gives the new instructor a safe place to practice new ways of interacting in the classroom and minimizes the risks of problems arising while switching cultures. In addition, since both instructors are listed as the instructor of record, neither can be held fully responsible for any negative student evaluations [40]. In terms of required time, during the co-teaching semester, the new instructor has the benefit of previously-used materials and only has to prepare for being in charge of the class about half the time. This leaves additional time and energy available for some of the other more reflective aspects of co-teaching. In addition, when teaching on his own the following semester, it was hoped that the experience with the course and a set of classroom instructional materials, old tests, and online resources that CH had previously compiled would make it less time consuming for MF to continue in a PhysTEC manner than a more traditional manner.

**D. Cost of Co-Teaching**

In the co-teaching model described in this paper, the only cost is that of a replacement to teach one class. This allows two instructors to co-teach a single class and have time for additional discussion and reflection without increasing the total amount of time spent on teaching duties. In the case of this project, an adjunct was hired using external grant money. The recommended part-time rate for a 4-credit class at WMU in Fall 2005 was $2,800. This rate is comparable to

national part-time faculty salary data [41]. Some departments, of course, may be able to absorb an extra class for one semester, thus, reducing the required cost to essentially nothing.

**4. Data Collection and Analysis**

The goal of this study was to develop a better understanding of the prospects of co-teaching for promoting instructional change through the in-depth investigation of one semester of co-teaching. This is what Stake refers to as an *instrumental case study* [42]. The expectation is that a deep understanding of this single case can be used to provide insight into the use of co-teaching in other similar settings.

**A. The Case**

The case under investigation was the Fall 2005 co-teaching of the first semester introductory calculus-based physics course at WMU by two of the authors – CH and MF. CH was an experienced instructor in his fourth year teaching at WMU. As a member of the WMU PhysTEC team he had been involved in the reform of the introductory calc-based physics sequence at WMU and had previously taught both of the courses in that sequence using reformed methods. He was also an experienced PER researcher with knowledge about many PER instructional interventions. MF was a new tenure-track faculty member in his first semester at WMU. All of his prior teaching experience was as a physics teaching assistant while a graduate student at the Ohio State University (OSU). As a graduate student at OSU he had some exposure to PER via his interactions with the OSU physics education research group who ran a required quarter-long course for the TAs. The purpose of co-teaching was to allow MF to gain enough experience with the WMU PhysTEC reforms that he would implement the PhysTEC principles in subsequent semesters. He was scheduled to teach the same course on his own in Spring 2006.

**B. Data Sources**

Case study research relies on multiple sources of evidence [43]. A faculty member from the college of education (AB) was asked to participate in the co-teaching experience as an outsider. She conducted open-ended individual interviews with CH and MF at the beginning, middle, and end of the co-teaching semester. Interviews focused on thoughts about how the course was going, general beliefs about teaching and learning, and perceptions about how the co-teaching was going. A final interview was conducted with MF at the end of Spring 2006 once he had taught the course on his own. The associated data sources were the interview transcripts of the seven, 45-75 min interviews. AB also observed both CH and MF teaching once at the beginning, middle, and end of the co-teaching semester. The associated data source was the field notes taken during each of the six observations scaffolded by the RTOP course observation form [44][45]. The final data source was the syllabi used by CH and MF in Fall 2005 for co-teaching and by MF in Spring 2006 when teaching alone as well as self-reported course structure for each semester.

**C. Data Analysis**

Both CH and AB independently analyzed all of the data sources looking for four things: 1) evidence related to MF's instructional practices; 2) evidence related to MF's beliefs about teaching and learning; 3) evidence related to MF's intentions towards future instruction; and 4) any other evidence related to co-teaching that seemed helpful in understanding the experience. After completing this independent analysis, CH and AB compared notes. There was a large degree of agreement between the two analyses and all disagreement was resolved through discussion.

**D. Credibility**

Efforts were made to ensure that this study provides an authentic portrait of the co-teaching experience [46]. As Yin suggests, multiple sources of evidence were used and all results are based on the complete set of evidence [43]. The analysis also involved two distinct perspectives – that of an insider in the experience (CH) and that of an outsider (AB). Although not directly involved in the data analysis, the other insider (MF) was asked to review and comment on the results of this study. All of the results reported in this paper have been agreed upon by all three authors from their unique perspectives.

**5. Results**

Our goal was to document changes, if any, and degree of agreement with PhysTEC design principles in MF's i) teaching practices, ii) beliefs about teaching and learning and iii) intentions towards future instruction. Each of these aspects is discussed separately below.

**A. Teaching Practices**

Observation data was used to document MF's teaching practices. AB observed both MF and CH teaching three class sessions spread throughout the co-teaching semester. Observations were scaffolded by the RTOP observation tool and, in addition, AB took approximately four pages of handwritten field notes during each observation to describe her perception of what was happening.

Both MF and CH received similar scores on the RTOP instrument for each class session as well as similar scores to one another [47]. This suggests that they were both working appropriately within the interactive class structure. AB did notice some more subtle differences, though. For example, in her first observation of MF she writes "*MF was somewhat more structured than I saw CH to be, but very interactive with students nonetheless. MF presented concepts and then problems that exemplified them. Less of having students generate concepts. More formulas.*"

MF also noticed this small difference [48].

>  *MF: I noticed CH's technique [for managing class discussions] is even slightly different from mine.*

>  *AB: How so?*

>  *MF: I am not criticizing him at all because this is his technique and it obviously works, but from my point of view, he doesn't mind letting the students hang for a long time and squirm and sweat over this problem. He will ask some, what I consider, very open ended questions whereas I will tend to ask something that I consider slightly more leading. (MF2#49-55)*

Co-Teaching    Page 15 of 30

Although noticeable, these differences were considered by AB to be minor compared to the large difference between the co-taught course and other science courses with which she was familiar. There were also no noticeable changes in MF's instructional practices during the semester. These observations suggest that the scaffolding provided by the course structure was effective, right from the start, in helping MF teach in a non-traditional way. There was no "implementation dip" present in MF's teaching performance.

While observations did not indicate any shift in instructional practices, MF perceived a shift in his own instruction towards more focus on concepts and less on mathematics. "*As the semester wore on what I ended up getting in the habit of . . . going through the concepts setting up the problem and saying to the students 'you go figure out the algebra on your own'. That allows you to go through many more problems and it also allows you to spend a larger percentage of time on the physics per problem so that they realize that the problem isn't a massive algebraic equation, but it really is physics.*" (MF3#262-267)

Without this structure it is likely that MF, much like the new faculty interviewed in Boice's study, would have put much more emphasis on facts and principles lecturing. The likelihood of this was confirmed by MF during the first interview (conducted during the first week of classes when MF had participated while CH was in charge of the class, but had not yet been in charge himself).

> *AB: If you were doing this by yourself, if they just say "okay, here is your class schedule for the semester. Good luck." What would you be doing? How would you approach preparing for a class like that?*

> *MF: I would probably not actually in all honesty . . . not have done it the same way that we are treating this class.*

> *AB: Because of?*

Co-Teaching                                                                                                               Page 16 of 30

*MF: I will probably treat it more like a lecture. Of course I tend to be more interactive, so I will still be more interactive, asking the students questions and things. I probably wouldn't do as many in-class activities as we are doing now. . . . and so it will probably be a little bit more like the formal lecture."* (MF1#222-233)

This was confirmed again with a similar question during the final interview with MF after the end of Spring 2006.

**B. Beliefs About Teaching and Learning**

MF's beliefs about how students learn appeared to be consistent throughout the semester and largely aligned with the beliefs behind the PhysTEC course structure. As a physics graduate student at Ohio State University, MF received TA training from the physics education research group that MF described as emphasizing the Socratic method and group work. He described developing favorable opinions of both. Thus, even though MF envisioned his teaching as some variation of a traditional lecture, he did not think that students would get a lot out of such a lecture. *"A student sitting in a lecture listening to you is going to do what most students do, and that is, fall asleep or walk away and not learn something. . . . You have to lead the student to an understanding by asking him questions."* (MF1#128-132). This disconnect between beliefs and practices appears to be common in all types of faculty [12, 49, 50].

Although his beliefs about student learning were consistent throughout the semester, his beliefs about teaching appeared to change. Although he envisioned his teaching as being more interactive than a traditional lecture, he was initially concerned about the PhysTEC course structure as being too much of a departure from the lecture method. *"I have really come to appreciate the use of in-class problems. It's surprising to know, because when I first came I was skeptical about having students do nothing but problems in class – just sort of standing by while they do problems. It really seems to be a good method."* (MF2#84-88) His largest initial concern appeared to be student resistance to such an interactive class structure. Thus he did not envision such methods being successful until he experienced the students being engaged and was also convinced by a survey of student perceptions of what helped them learn. *"What convinced*

Co-Teaching                                                                                                                                                      Page 17 of 30

*me about this [the PhysTEC course structure] was that most of the students . . . were really engaged . . . but even more than that at the end of the semester when we gave them the survey, the thing they liked the most was the in-class work. Very strangely surprisingly to me was that they liked doing this and found it to be very helpful to them."* [MF2#184-188]

**C. Intention Towards Future Instruction**

Not surprisingly, as MF's beliefs about teaching changed, his intentions towards future instruction also began to change. From the first three interviews, it appeared that his intentions towards future instruction, specifically the following semester, were changing to become more aligned with a PhysTEC style course. As noted earlier, MF was initially skeptical about the course design where much of the class time was used having students work together on problems. By the mid-term interview, MF was beginning to become comfortable with the course design, but was still largely non-committal about how these might fit into his future instruction. *"You know, it [the co-teaching experience] taught me something that I am going to adopt aspects of in future courses."* [MF2#196-197] By the end of term, though, he seemed to have shifted his perception to be very favorable towards the course structure. *"My class [next semester] is going to be very similar to what we did last semester, even the structure will be the same structure. It's going to be almost identical."* (MF3#272-273)

As discussed earlier, one of the things that led to a change of intentions at the end of the course was the students' unexpected (to MF) positive response to the course. This concern for student opinion was actually a theme that ran throughout all of the interviews and is consistent with Boice's finding that new faculty tend to "teach defensively, so as to avoid public failures at teaching" (ref [35], p. 170).

We can also look at the changes made to the initial course and reasons for the changes. Even though at the end of the Fall co-teaching experience, MF indicated that his Spring 2006 course would be "almost identical" to the co-taught course, he later decided to make some changes to the course structure. A comparison of the Fall 2005 (co-teaching) and Spring 2006 (MF teaching alone) course structure can be found in Tables 2 and 4. Although the Spring 2006 course appears to be well within the PhysTEC course structure, all movement was towards a more traditional course structure. In addition, almost all changes were made in order to reduce faculty time required or to reduce perceived student dissatisfaction. For example, in Spring 2006, MF decided to change the written homework from a group to an individual assignment. This was largely based on his perception that students did not like the group homework assignments. "*It [changing to individual homework] takes out complaints such as' he doesn't do his share of the work.*'" [MF3#99] In contrast, although quiz corrections were quite popular with students, MF decided not to use quiz corrections in Spring 2006 due primarily to the extra time required for grading.

The final interview at the conclusion of the Spring 2006 semester revealed that MF was unhappy with many of the changes that he made and planned to go back to a course structure more closely aligned to the Fall 2005 course. He indicated that his direct experience with co-teaching followed by teaching alone convinced him that the course elements were important enough in promoting student learning that they were worth extra time and possible student dissatisfaction. "*I did not do quiz corrections this year, simply because of time constraints involved, and, looking back on that, I think that was a bad idea. . . . I think students looked at quizzes as sort of a module of the course and once you are done with the quiz you are done with learning that material.*" (MF4#164-168). "*I'm going to readopt those [quiz corrections and group homework] and, it's going to be extra time involved, but in my mind it's worth it.*" (MF4#182-184)

## 6. Discussion

Co-teaching appears to have been successful in changing MF's beliefs and intentions towards instructional practices consistent with the WMU PhystTEC design principles. There did not appear to be any changes in instructional practices during the semester, perhaps due to the predetermined course structure that constrained possible practices. Beliefs about teaching and learning appeared to be largely aligned with the PhysTEC design principles by the middle of the semester, while plans for future teaching appeared to continue to change throughout the co-teaching semester and ended largely PhysTEC compatible. Thus, it appears the entire co-teaching semester was important. It probably would not have been enough, for example, to just co-teach for the first half of the semester.

We conclude that there were three important components to the co-teaching design: 1) it lasted an entire semester, 2) the course structure was set up in advance by the experienced instructor, and 3) there was a collegial, cooperative relationship between the co-teachers. It was not a student-teacher or mentor-mentee type of relationship. *"Well the thing that I liked the most about this is it wasn't like I was Charles' protégé. He recognizes me as a colleague and we were teaching this class together. . . . it wasn't like teacher-apprenticeship which at this level it might seem sort of insulting."* (F3#283-286)

An unexpected, yet valuable outcome of co-teaching was that informal non teaching related discussions helped MF become acculturated to WMU in areas other than teaching. *"I would ask him [CH] everything, not just about teaching. . . . He was actually very helpful in a lot of areas including grant writing."* (MF3#398-401). *"These discussions often sprang from side conversations during the first five minutes before class while waiting for people to mingle in the class."* (MF3#423-424) As Boice noted, this additional support is frequently lacking for new faculty [35].

Of course, this study has many limitations that caution against generalization and suggest further work is needed. We note that this study was of a single case. More examples of similar cases are needed. In addition, MF began co-teaching with favorable views of research-compatible instructional practices. It is not clear that this model would be successful with a new faculty member hostile to research-based instruction. It is also likely that the personalities and other personal characteristics of the co-teaching participants must be somewhat compatible for co-teaching to be effective.

Other interesting questions for future work would be to examine the applicability of co-teaching for other populations. MF was a new faculty member. Would co-teaching work similarly with an experienced faculty member who had already established a strong traditional teaching routine? Would the co-teaching model be a useful way for graduate students to develop teaching expertise [51]?

## 7. Conclusions

Co-teaching is a cost-effective model that shows significant promise as an effective way to promote research-consistent instruction in new faculty. It appears to be able to provide a much larger change for much less money than common workshop models of dissemination since the only cost is cost of a replacement (adjunct) to teach one course so two instructors can team up without increasing their workload. It is effective because it immerses the new instructor in a teaching role in the new instructional context and provides scaffolding and modeling to ensure success.

Of course, co-teaching is only appropriate when there is a teacher available who is experienced in teaching the target course in a research-consistent manner. Thus, co-teaching cannot solve all dissemination problems. Yet, when the conditions are right, it should not be overlooked. With the expanding presence of PER researchers/teachers in physics departments, it could have a significant impact.

## 8. Acknowledgements

This project was supported, in part, by the Physics Teacher Education Coalition (PhysTEC), funded by the National Science Foundation and jointly administered by the American Physical Society, American Association of Physics Teachers, and the American Institute of Physics. The authors would like to thank Alvin Rosenthal for his helpful comments on an earlier version of this manuscript.

Table 1: Design Principles of WMU PhysTEC Courses (Departures from traditional instruction)

1. Students should be actively engaged with the material during class time. This is best accomplished via student-student interaction.
2. Students should read the text before coming to class and most will not do this unless there is some sort of enforcement.
3. Class discussions and tests should place significant emphasis on conceptual issues and qualitative questions.
4. Class discussions and tests should place significant emphasis on the solving of multi-step problems (i.e., ones that cannot be solved by substituting numbers into a single equation).
5. Student problem solutions should start from basic principles and contain written explanation of reasoning.
6. Test questions should require students to engage in the desired thinking processes. This means that test questions should not be similar enough to questions students have previously seen that a rote strategy is fruitful.
7. Formative assessment, both informal and formal, should be used to determine students' current understanding for the purpose of designing appropriate subsequent instruction.
8. Depth of student understanding should be valued more than breadth of content covered during the course.

Table 2: Significant course elements of Physics 2050 during Fall 2005 (co-teaching) and Spring 2006 (MF teaching alone).

| Course Component | Course Component – CH and MF course (Fall 2005) | Course Component – MF course (Spring 2006) |
| --- | --- | --- |
| Encourage students to read the text before class | Reading Assignment – students asked to submit a question about the assigned reading [52] as well as to discuss an example problem from the reading assignment [53]. Typically due Monday evening – submitted via WebCT. | Reading Quiz – WebCT quiz covering important aspects of the reading. Typically due Monday evening. [Approximately 9 questions/week. Similar to CH/MF online exercises – highest of 4 chances] |
| Use of class time | During class time there may be short lectures (5-10 minutes) when new topics are introduced. But, most of the class time is spent with students working in assigned groups (and, usually white boards) on instructor-assigned activities. Activities varied from conceptual questions to quantitative problems – all were designed or chosen to encourage discussion and thought about the physics topics and discourage rote learning. The instructor(s) moved around and interacted with groups. When the instructor noted that many of the groups were finished or stuck, he would hold a whole-class discussion. This would start by calling on a group (typically at random, but sometimes chosen based on observations). | |
| Assignment of groups | At the beginning of the semester, students are assigned to groups of 3-4 based on where they live (as homogeneous as possible) and their performance in the prerequisite math course and conceptual pretest (as heterogeneous as possible). Gender was also considered, but given less weight than the first two factors. Students remained in these groups throughout the semester. | |
| Online Exercises | Each week 6-12 exercises were assigned. These consisted of conceptually-oriented questions (often multiple-choice) or relatively simple calculations and were typically due on Wednesdays at class time. Grade recorded is average of up to 4 chances. | Used for reading quizzes |
| Online Homework | Students were each responsible for completing their individual assignments (although they were explicitly encouraged to work together). Each week 4 problems were assigned. These were similar to mid-level problems found in standard textbooks (they involved multiple steps and were specifically chosen to not be easily solvable by rote) and were typically due on Thursdays at class time. Grade recorded is average of up to 4 chances. | Online HW was optional |
| Written Homework | Each group of students was responsible for turning in a written solution (using the problem solving format) to each of the 4 WebCT problems. | Each individual student was responsible for turning in a written solution for 6-10 HW problems. Problems were similar (or identical) to the CH/MF WebCT problems. |
| Problem Solving Framework | Students required to include at least three things in their problem solutions: General Approach (big picture description of how to solve the problem, including relevant physics principles), Procedure (details about how to solve the problem, including specific steps), Implementation (working out the details and evaluating the result). This framework was modeled by instructor during class. | Problem solving framework was emphasized in class but not required in student solutions. Student solutions were expected to start from basic principles and show reasoning. |

| Use of Main Ideas | The important physics concepts covered during the course were broken into 21 main ideas that explicitly categorized, frequently referred to in class and provided to students on exams. All problem solutions were expected to be based on one or more main idea. | |
|---|---|---|
| Testing | Each week a quiz or exam was given. The weekly quizzes focused on the material covered during the week. Every third week, there was an exam that focused on the material covered during the previous three weeks. Each quiz and exam had a similar format. There were 2-4 conceptually-oriented short answer questions and one multi-step quantitative problem that required a solution using the problem solving format. There was also a final exam with 3 multi-part conceptual questions and 3 problems. Complete written solutions were available on WebCT shortly after each examination. | |
| Quiz Corrections | After graded quizzes were returned, students had the option to complete a quiz correction assignment. The purpose of the assignment was to have students reflect on their quiz mistakes in light of the instructor solution and construct generalized physics knowledge. Quiz correction attempts resulted in an increased quiz grade of 50% of the lost points if done well and no change in the quiz grade if done poorly. | Not used |

Table 3: Alignment of co-teaching activities within the cognitive apprenticeship framework.

| | Modeling | Coaching | Scaffolding | Articulation | Reflection | Exploration |
|---|---|---|---|---|---|---|
| (1) CH and MF alternate being in charge of class each week | X | X | | | | X |
| (2) Weekly meetings between CH and MF to reflect on previous week and discuss initial plans for coming week. | X | X | | X | X | |
| (3) Course structure set up by CH to support PhysTEC design principles | | | X | | | |
| (4) MF had access to materials used by CH in previous offerings of the course | | | X | | | |
| (5) MF teaches course on his own during subsequent semester | | | | | | X |

Table 3: Information from course syllabi describing contribution of various course elements towards the final course grade

| Component | CH and MF course (Fall 2005) | MF course (Spring 2006) |
|---|---|---|
| Exam Average | 40% (4 exams) | 30% (3 exams) |
| Final Exam (comprehensive) | 20% | 20% |
| Quiz Average | 15% | 15% |
| Online Problems | 5% | -- |
| Written Problems | 5% (group) | 20% (individual) |
| Online Exercises | 5% | -- |
| Reading Assignment | 5% (reading questions) | 10% (reading quiz) |
| In-Class Group Work (all members get same score) | 5% | 5% |
| Total | 100% | 100% |